\documentclass[aps,prb,twocolumn,prabib,amsmath,amssymb,superscriptaddress,notitlepage]{revtex4-1}
\usepackage{graphics}
\usepackage{bm}
\usepackage{dcolumn}
\usepackage{natbib}
\usepackage{epstopdf}
\usepackage{float}
\usepackage{graphicx}
\usepackage{epsfig}
\usepackage[pdfstartview=FitH]{hyperref}
\usepackage{color}
\usepackage{appendix}
\usepackage{ulem}

\newcommand{\rr}{{\bf r}}

\begin{document}
\title{Two-component quantum Hall effects in topological flat bands}
\author{Tian-Sheng Zeng}
\affiliation{Department of Physics and Astronomy, California State University, Northridge, California 91330, USA}
\author{W. Zhu}
\affiliation{Theoretical Division, T-4 and CNLS, Los Alamos National Laboratory, Los Alamos, New Mexico 87545, USA}
\author{D. N. Sheng}
\affiliation{Department of Physics and Astronomy, California State University, Northridge, California 91330, USA}
\date{\today}
\begin{abstract}
We study quantum Hall states for two-component particles (hardcore bosons and fermions) loading in topological lattice models. By tuning the interplay of interspecies and intraspecies interactions, we demonstrate that two-component fractional quantum Hall states emerge at certain fractional filling factors $\nu=1/2$ for fermions ($\nu=2/3$ for bosons) in the lowest Chern band, classified by features from ground states including the unique Chern number matrix (inverse of $\mathbf{K}$-matrix), the fractional charge and spin pumpings, and two parallel propagating edge modes. Moreover, we also apply our strategy to two-component fermions at integer filling factor $\nu=2$, where a possible topological Neel antiferromagnetic phase is under intense debate very recently. For the typical $\pi$-flux checkerboard lattice, by tuning the onsite Hubbard repulsion, we establish a first-order phase transition directly from a two-component fermionic $\nu=2$ quantum Hall state at weak interaction to a topologically trivial antiferromagnetic insulator at strong interaction, and therefore exclude the possibility of an intermediate topological phase for our system.
\end{abstract}
\maketitle
\section{introduction}
Fractionalized topological ordered phases in topological flat bands have attracted intense attention in the past few years~\cite{BL2013}. They emerge as the ground state of interacting many-body systems at the fractional fillings of topological bands in the absence of a magnetic field, in analogy to the fractional quantum Hall (FQH) states in two-dimensional Landau levels~\cite{Sun2011,Neupert2011,Sheng2011,Tang2011}. Topological flat bands with higher Chern number $C$ have been revealed  to host a series of Abelian single-component FQH states emerge at fillings $\nu=1/(C+1)$ (for hardcore bosons) and $\nu=1/(2C+1)$ (for spinless fermions)~\cite{LBFL2012,Wang2012r,Yang2012,Sterdyniak2013,Wang2013,Repellin2014,Yao2015,Moller2015}, which are believed to be color-entangled lattice versions of multicomponent Halperin $(mmn)$ FQH states~\cite{Halperin1983} in a single Chern band~\cite{Barkeshli2012,Wu2013,Wu2014}. Much stronger evidence comes from a hidden symmetry of particle entanglement spectrum, as discussed in Ref.~\cite{Sterdyniak2013}. However, a closely related problem is that no direct evidence for the integer valued symmetric $\mathbf{K}$-matrix has been revealed, which is believed to classify the topological order at different fillings for multicomponent systems~\cite{Wen1992a,Wen1992b,Blok1990,Blok1991}.

Here we numerically address the possibility of multicomponent quantum Hall states in topological flat band models filled with interacting two-component (or bilayer) particles, where ``two-component'' serves as a generic label for spin or pseudo-spin (bilayer, etc.) quantum number. Different from the single-component case, there are more tunable parameters in two-component systems, like interspecies interaction whose magnitude can be tuned through Feshbach resonance in cold atom setting~\cite{Chin2010}. When the spin degrees of freedom are included, one would  expect many more exotic  phases to occur, such as quantum Hall ferromagnetism~\cite{Neupert2012,Kumar2014,Grushin2014}, and a rich class of fractional quantum Hall states including Halperin $(331)$ states for two-component fermions~\cite{Yoshioka1988,Yoshioka1989,He1991,He1993,Peterson2010,Thiebau2015} and spin-singlet incompressible states including Halperin $(221)$ states for two-component bosons~\cite{Ardonne1999,Furukawa2012,YHWu2013,Grass2014,Grass2012} previously studied for the lowest Landau level systems. Thus it is interesting to study the effect of interspecies interactions on two-component quantum Hall states in topological lattice models. Experimentally, the Haldane honeycomb insulator has been achieved from two-component fermionic $^{40}$K atoms with a tunable Hubbard repulsion in a periodically modulated honeycomb optical lattice~\cite{Jotzu2014}. For two-component bosonic atoms $^{87}$Rb in different hyperfine spin channels, the two-dimensional Hofstadter-Harper Hamiltonian is also engineered in the optical lattice with time-reversal symmetry~\cite{Aidelsburger2011}. These advances would open up new possibilities of studying the multicomponent quantum Hall effect for bosons and fermions in topological lattice models.

The aim of this paper is to provide compelling numerical evidence of $\mathbf{K}$ matrix classifications for both fractional Halperin and integer quantum Hall states in several microscopic topological lattice models through exact diagonalization (ED) and density-matrix renormalization group (DMRG) methods. By tuning the interspecies and intraspecies interactions, we show that for a given fractional filling factor, the many-body ground states in the decoupled limit evolve to a set of degenerate states separated from the higher-energy spectrum by a finite gap in the strongly interacting regime, whose topological nature is described by the $\mathbf{K}$-matrix~\cite{Blok1990}. In addition, the topological properties of these states are also characterized by (\textrm{i}) fractional quantized topological invariants related to Hall conductance, and (\textrm{ii}) degenerate ground states manifold under the adiabatic insertion of flux quanta. For integer quantum Hall states, we mainly focus on their phase transition nature driven by onsite interspecies interactions, and demonstrate their first-order characteristics from discontinuous behaviors of related physical quantities at the transition point.

This paper is organized as follows. In Sec.~\ref{model}, we give a description of the Hamiltonian of two-component quantum particles in two types of topological lattice models, such as $\pi$-flux checkerboard and Haldane-honeycomb lattices. In Sec.~\ref{ground}, we study the ground states of these two-component particles in the strong interaction regime, present numerical results of the $\mathbf{K}$ matrix by exact diagonalization at fillings $\nu=2/3$ for spinful hardcore bosons and $\nu=1/2$ for spinful fermions in Sec.~\ref{kmatrix}, and discuss the properties of these ground states under the insertion of flux quanta. In Sec.~\ref{pumping}, we calculate the adiabatic charge and spin pumping from DMRG, and demonstrate the quantized drag Hall conductance. In Sec.~\ref{edge}, we discuss the momentum-resolved entanglement spectrum of these two-component FQH states. In Sec.~\ref{iqh}, we discuss the role of interspecies onsite interaction on the $C_q=2$ integer quantum Hall state for two-component fermions at $\nu=2$. Finally, in Sec.~\ref{summary}, we summarize our results and discuss the prospect of investigating nontrivial topological states in multicomponent quantum gases.

\section{The model Hamiltonian}\label{model}
Our starting point is the following noninteracting Hamiltonian of two-component particles (hardcore bosons and fermions) in topological lattice models, such as the $\pi$-flux checkerboard (CB) lattice,
\begin{align}
  &H_{CB}=-t\sum_{\sigma}\!\!\sum_{\langle\rr,\rr'\rangle}\!\big[c_{\rr',\sigma}^{\dag}c_{\rr,\sigma}\exp(i\phi_{\rr'\rr})+H.c.\big]\nonumber\\
  &\pm t'\!\sum_{\sigma}\!\!\!\sum_{\langle\langle\rr,\rr'\rangle\rangle}\!\!\!\! c_{\rr',\sigma}^{\dag}c_{\rr,\sigma}
  -t''\!\sum_{\sigma}\!\!\!\sum_{\langle\langle\rr,\rr'\rangle\rangle}\!\!\!\! c_{\rr',\sigma}^{\dag}c_{\rr,\sigma}+H.c.,
\end{align}
and Haldane-honeycomb (HC) lattice
\begin{align}
  &H_{HC}=-t'\sum_{\sigma}\sum_{\langle\langle\rr,\rr'\rangle\rangle}[c_{\rr',\sigma}^{\dag}c_{\rr,\sigma}\exp(i\phi_{\rr'\rr})+H.c.]\nonumber\\
  &-t\!\sum_{\sigma}\!\!\!\sum_{\langle\rr,\rr'\rangle}\!\!\!\! c_{\rr',\sigma}^{\dag}c_{\rr,\sigma}
  -t''\!\sum_{\sigma}\!\!\!\sum_{\langle\langle\rr,\rr'\rangle\rangle}\!\!\!\! c_{\rr',\sigma}^{\dag}c_{\rr,\sigma}+H.c.,
\end{align}
where $c_{\rr,\sigma}^{\dag}$ is the particle creation operator of spin $\sigma=\uparrow,\downarrow$ at site $\rr$, $\langle\ldots\rangle$,$\langle\langle\ldots\rangle\rangle$ and $\langle\langle\langle\ldots\rangle\rangle\rangle$ denote the nearest-neighbor, the next-nearest-neighbor, and the next-next-nearest-neighbor pairs of sites, respectively. We take the parameters $t'=0.3t,t''=-0.2t$ for checkerboard lattice, as in Ref.~\cite{Zeng2015}, while $t'=0.6t,t''=-0.58t$ for honeycomb lattice, as in Refs.~\cite{Wang2011,Wang2012}. These parameters enhance the flatness of the topological band and make quantum Hall states more robust.

Consider the on-site interspecies and nearest neighboring intraspecies interactions
\begin{align}
  V_{int}=U\sum_{\rr}n_{\rr,\uparrow}n_{\rr,\downarrow}+V\sum_{\sigma}\sum_{\langle\rr,\rr'\rangle}n_{\rr',\sigma}n_{\rr,\sigma}
\end{align}
where $n_{\rr,\sigma}$ is the particle number operator of spin-$\sigma$ at site $\rr$. The model Hamiltonian becomes $H=H_{CB}+V_{int}$ ($H=H_{HC}+V_{int}$). Here, $U$ is the strength of the interspecies interaction while $V$ is the strength of intraspecies correlations in topological flat bands, playing the analogous role of Haldane pseudopotentials for two-component FQHE system in Landau levels~\cite{Peterson2010}. For strong Hubbard repulsion, two-component fermions in the Haldane-honeycomb model may exhibit various chiral magnetic orderings or topological Mott insulator phase at half-filling~\cite{Hickey2016,Wu2016}.

In the ED study, we explore the many-body ground state of $H$ in a finite system of $N_x\times N_y$ unit cells (the total number of sites is $N_s=2\times N_x\times N_y$). The total filling of the lowest Chern band is $\nu=2N/N_s$, where $N=N_{\uparrow}+N_{\downarrow}$ is the total particle number with global conservation $U(1)$-symmetry. With the translational symmetry, the energy states are labeled by the total momentum $K=(K_x,K_y)$ in units of $(2\pi/N_x,2\pi/N_y)$ in the Brillouin zone. For larger systems, we exploit DMRG on cylinder geometry, and keep the number of states 1200--2400 to obtain accurate results.

\section{Halperin (221) and (331) states}\label{ground}
In this section, we present the numerical evidences of two-component FQH states at filling factors $\nu=2/3$ and $\nu=1/2$ for bosons and fermions, respectively. Importantly, the Chern number matrix (the inverse of $\mathbf{K}$-matrix) uniquely identifies these two-component FQH states. The further information from ground state degeneracies and charge (spin) pumpings is complementary to and consistent with the Chern number matrix. The Chern number matrix also provides an accurate prediction for the transport measurement for experimental systems. Thus it is very important to numerically extract this topological information for two-component particles in topological flat band models.

\begin{figure}[t]
  \includegraphics[height=1.8in,width=3.4in]{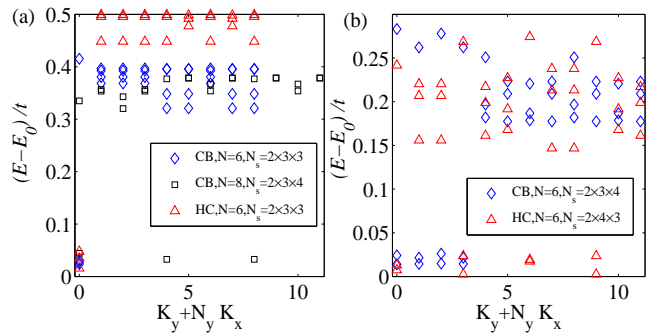}
  \caption{\label{energy}(Color online) Numerical results for the low energy spectrum of (a) bosonic systems $\nu=2/3,N_s=3\times N$ in different topological lattices at $U=10t,V=0$ and (b) fermionic systems $\nu=1/2,N_s=N\times4$ in different topological lattices at $U=10t,V=20t$.}
\end{figure}

\subsection{Ground-state degeneracy}

First, we demonstrate the ground state degeneracy on torus geometry, which serves as a primary signature of an incompressible FQH state. We consider finite-size systems up to maximum particle number $N=8$. In Figs.~\ref{energy}(a) and~\ref{energy}(b), we show  the energy spectrum of several typical systems in strong interacting regime. The key feature is that, there exists a well-defined and degenerate ground state manifold separated from higher-energy levels by a robust gap.
For two-component hardcore bosons at $\nu=2/3$, the ground states show three-fold quasi-degeneracies; For two-component fermions at $\nu=1/2$, we find that the ground state manifold hosts eight-fold quasi-degeneracies. We also calculate the density and spin structure factors for the ground states, and exclude any possible charge or spin density wave orders as the competing ground states, due to the absence of the Bragg peaks in the results (We check them up to $N_s=2\times4\times4$ sites using DMRG with periodic boundary conditions).

\begin{figure}[t]
  \includegraphics[height=1.7in,width=3.4in]{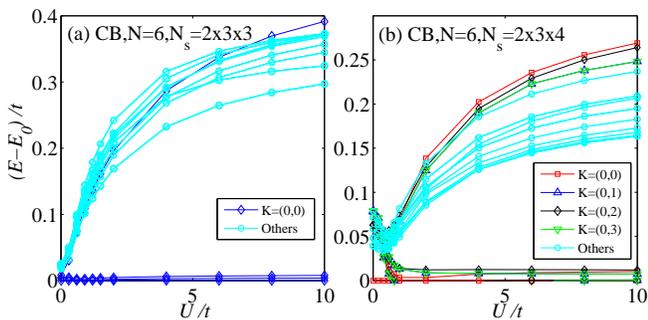}
  \caption{\label{phaseU}(Color online) Robustness of degenerate ground states on the $\pi$-flux checkerboard lattice vs the variations of interspecies repulsion for: (a) bosonic systems $N=6,N_s=2\times3\times3$ at $\nu=2/3,V=0$; (b) fermionic systems $N=6,N_s=2\times3\times4$ at $\nu=1/2,V=20t$. The energies are measured relative to the lowest ground energy. The momentum sectors in which these degenerate FQH states emerge are labeled explicitly with different colors while the other momentum sectors are labeled by the cyan circle.}
\end{figure}

Next, we consider the effects of interspecies repulsion on these topological phases. We first use ED to study the evolution of the energy spectrum with $U$ for the periodic system. As shown in Figs.~\ref{phaseU}(a) and ~\ref{phaseU}(b), with the decrease of $U$, the degeneracy  is lifted and finally disappears at $U=0$, where the ground state becomes a possibly metallic phase with vanishingly small excitation energy gap. In usual FQH states, the occupation of each single-particle orbital is constant and equal to the filling factor~\cite{Regnault2011}. By diagonalizing the $N_s\times N_s$-matrix $\rho_{\sigma}=\langle c_{\rr',\sigma}^{\dag}c_{\rr,\sigma}\rangle$, we obtain reduced single-particle eigenstates $\rho_{\sigma}|\phi_{\alpha}\rangle=\rho_{\sigma}^{\alpha}|\phi_{\alpha}\rangle$ where $|\phi_{\alpha}\rangle$ ($\alpha=1,\ldots,N_s$) are the effective orbitals as eigenvectors for $\rho_{\sigma}$ and $\rho_{\sigma}^{\alpha}$ ($\rho_{\sigma}^1\geq\ldots\geq \rho_{\sigma}^{N_s}$) are interpreted as occupations. We find that the occupations are close to the uniform filling $\rho_{\sigma}^{\alpha}\simeq\nu/2$ for $\alpha\leq N_s/2$ with standard deviations of the order $0.01$, while $\rho_{\alpha}\ll1$ for $\alpha>N_s/2$ in the strongly interacting regime, indicating an incompressible liquid with particles uniformly occupying the $N_s/2$-orbitals of the lowest Chern band. However, for $U\ll t$, these  eigenvalues  form  a nonuniform distributions with standard deviations of the order $0.2$, which are possibly compressible liquid states. Thus, we remark that the emergence of two-component FQH states is induced by suitable interspecies and intraspecies repulsions.

\subsection{Chern number matrix and $\mathbf{K}$-matrix}\label{kmatrix}

\begin{figure}[t]
  \includegraphics[height=1.8in,width=3.4in]{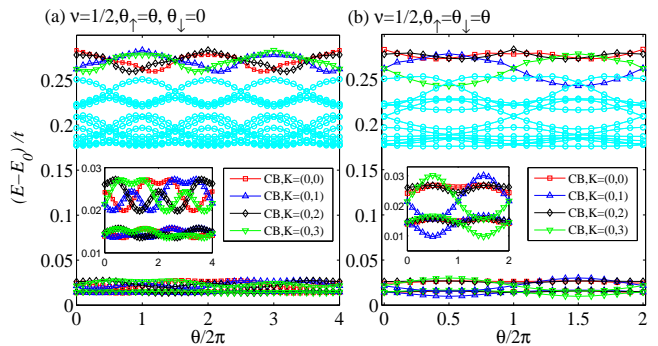}
  \caption{\label{flux}(Color online) Numerical results for the y-direction spectral flow of fermionic systems $N=6,N_s=2\times3\times4$ at $\nu=1/2,U=10t,V=20t$: (a) $\theta_{\uparrow}^{\alpha}=\theta,\theta_{\downarrow}^{\alpha}=0$; (b) $\theta_{\uparrow}^{\alpha}=\theta_{\downarrow}^{\alpha}=\theta$.}
\end{figure}

To uncover the topological nature of the ground-state manifold we extract the Chern number matrix. Here, we utilize the scheme proposed by one of the current authors in Refs.~\cite{Sheng2003,Sheng2006}.
With twisted boundary conditions $\psi(\rr_{\sigma}+N_{\alpha})=\psi(\rr_{\sigma})\exp(i\theta_{\sigma}^{\alpha})$ where $\theta_{\sigma}^{\alpha}$ is the twisted angle for spin-$\sigma$ particles in the direction-$\alpha$, we plot the low energy spectra flux under the variation of $\theta_{\sigma}^{\alpha}$. As shown in Figs.~\ref{flux}(a) and~\ref{flux}(b), these ground states evolve into each other without mixing with the higher levels. For two-species hardcore bosons at $\nu=2/3$, the energy recovers itself after the insertion of three flux quanta for $\theta_{\uparrow}^{\alpha}=\theta_{\downarrow}^{\alpha}=\theta$ or $\theta_{\uparrow}^{\alpha}=\theta,\theta_{\downarrow}^{\alpha}=0$, indicating its $1/3$ fractional quantization of quasiparticles.
Interestingly, for two-species fermions at $\nu=1/2$, the energy recovers itself after the insertion of two flux quanta for $\theta_{\uparrow}^{\alpha}=\theta_{\downarrow}^{\alpha}=\theta$, or after the insertion of four flux quanta for $\theta_{\uparrow}^{\alpha}=\theta,\theta_{\downarrow}^{\alpha}=0$, indicating its $1/4$ fractional quantization of quasiparticles. Meanwhile, the many-body Chern number of the ground state wavefunction $\psi_{i}$ is defined as~\cite{Sheng2003,Sheng2006}
\begin{align}
  C_{\sigma,\sigma'}^i=\frac{1}{2\pi}\int d\theta_{\sigma}^{x}d\theta_{\sigma'}^{y}\mathbf{Im}\left(\langle{\frac{\partial\psi^i}{\partial\theta_{\sigma}^x}}|{\frac{\partial\psi^i}{\partial\theta_{\sigma'}^y}}\rangle
-\langle{\frac{\partial\psi^i}{\partial\theta_{\sigma'}^y}}|{\frac{\partial\psi^i}{\partial\theta_{\sigma}^x}}\rangle\right)
\end{align}
For the three ground states with $K=(0,0)$ of two-species hardcore bosons at $N=6,N_s=2\times3\times3$, by numerically calculating the Berry curvatures using $m\times m$ mesh squares in the boundary phase space with $m\geq9$ we obtain $\sum_{i=1}^{3} C_{\uparrow,\uparrow}^i=2$, and $\sum_{i=1}^{3} C_{\uparrow,\downarrow}^i=-1$. Due to the symmetry $c_{\uparrow}\leftrightarrow c_{\downarrow}$, one has the symmetric properties $C_{\sigma,\sigma'}^i=C_{\sigma',\sigma}^i$. All of the above imply a symmetric $C$-matrix in the spanned  Hilbert space, namely,
\begin{align}
  \mathbf{C}=\begin{pmatrix}
C_{\uparrow,\uparrow} & C_{\uparrow,\downarrow}\\
C_{\downarrow,\uparrow} & C_{\downarrow,\downarrow}\\
\end{pmatrix}=\frac{1}{3}\begin{pmatrix}
2 & -1\\
-1 & 2\\
\end{pmatrix}
\end{align}
where the off-diagonal part $C_{\uparrow,\downarrow}$ is related to the drag Hall conductance. Thus we can obtain the $\mathbf{K}$-matrix which is the inverse of the $\mathbf{C}$-matrix, namely $\mathbf{K}=\mathbf{C}^{-1}=\begin{pmatrix}
2 & 1\\
1 & 2\\
\end{pmatrix}$. Therefore, we establish that three-fold ground states for two-species hardcore bosons at $\nu=1/3$ are indeed Halperin (221) states in the lattice version, and the three-fold degeneracy coincides with the determinant $\det\mathbf{K}$, as predicted in Ref.~\cite{Wen1995}.

Similarly, for the eight ground states with $K=(0,i), (i=0,1,2,3)$ of two-species fermions at $N=6,N_s=2\times3\times4$, by numerically calculating the Berry curvatures (using $m\times m$ mesh squares in boundary phase space with $m\geq9$), we obtain $\sum_{i=1}^{8} C_{\uparrow,\uparrow}^i=3$, and $\sum_{i=1}^{8} C_{\uparrow,\downarrow}^i=-1$. The above results imply a symmetric $C$-matrix, namely,
\begin{align}
  \mathbf{C}=\begin{pmatrix}
C_{\uparrow,\uparrow} & C_{\uparrow,\downarrow}\\
C_{\downarrow,\uparrow} & C_{\downarrow,\downarrow}\\
\end{pmatrix}=\frac{1}{8}\begin{pmatrix}
3 & -1\\
-1 & 3\\
\end{pmatrix}
\end{align}
Thus we can obtain the $\mathbf{K}$-matrix which is the inverse of the $C$-matrix, namely $\mathbf{K}=\mathbf{C}^{-1}=\begin{pmatrix}
3 & 1\\
1 & 3\\
\end{pmatrix}$. Therefore, we establish that eight-fold ground states for two-species fermion at $\nu=1/2$ are indeed Halperin (331) states in lattice version, and the eight-fold degeneracy coincides with the determinant $\det\mathbf{K}$, as predicted in Ref.~\cite{Wen1995}.

With $\theta_{\uparrow}^{x}=\theta_{\downarrow}^{x}=\theta^{x}$ and $\theta_{\uparrow}^{y}=\theta_{\downarrow}^{y}=\theta^{y}$, the many-body charge Chern number of the ground state wavefunction, related to charge Hall conductance, reads
\begin{align}
  C_{q}&=\int\frac{d\theta^{x}d\theta^{y}}{2\pi}\mathbf{Im}
  \left(\langle{\frac{\partial\psi}{\partial\theta^x}}|{\frac{\partial\psi}{\partial\theta^y}}\rangle
  -\langle{\frac{\partial\psi}{\partial\theta^y}}|{\frac{\partial\psi}{\partial\theta^x}}\rangle\right)\nonumber\\
&=\mathbf{q} \cdot \mathbf{C}\cdot \mathbf{q}^{T} =\sum_{\sigma,\sigma'}C_{\sigma,\sigma'}=\nu, \label{charge}
\end{align}
where $\mathbf{q}=(1,1)$ is the charge eigenvector of $\mathbf{K}$-matrix. Similarly, with $\theta_{\uparrow}^{x}=-\theta_{\downarrow}^{x}=\theta^{x}$ and $\theta_{\uparrow}^{y}=-\theta_{\downarrow}^{y}=\theta^{y}$, we can also define the many-body spin Chern number of the ground state wavefunction, related to spin Hall conductance, as
\begin{align}
  C_{s}&=\int\frac{d\theta^{x}d\theta^{y}}{2\pi}\mathbf{Im}
  \left(\langle{\frac{\partial\psi}{\partial\theta^x}}|{\frac{\partial\psi}{\partial\theta^y}}\rangle
-\langle{\frac{\partial\psi}{\partial\theta^y}}|{\frac{\partial\psi}{\partial\theta^x}}\rangle\right)\nonumber\\
&=\mathbf{s} \cdot \mathbf{C}\cdot \mathbf{s}^{T} =C_{\uparrow,\uparrow}+C_{\downarrow,\downarrow}-C_{\uparrow,\downarrow}-C_{\downarrow,\uparrow},  \label{spin}
\end{align}
where $\mathbf{s}=(1,-1)$ is the spin eigenvector of $\mathbf{K}$-matrix. From Eqs.~\ref{charge} and~\ref{spin}, we conclude that to identify the nature of these degenerate states, one can calculate the adiabatic charge and spin pumping by performing different flux insertion simulations, which identify the total fractionally quantized Hall and drag Hall conductances in experiments, as will be shown below in Sec.~\ref{pumping}.

\subsection{Fractional charge and spin pumpings}\label{pumping}

\begin{figure}[t]
  \includegraphics[height=1.75in,width=3.4in]{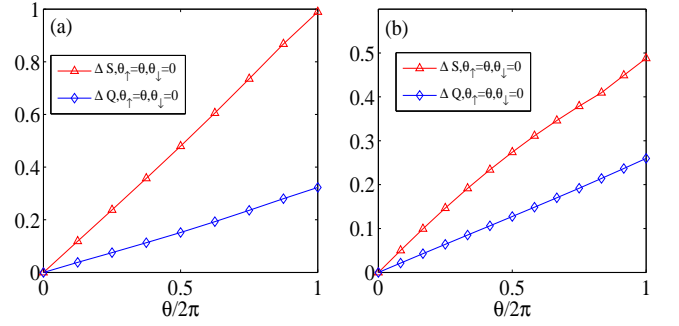}
  \caption{\label{pump}(Color online) The charge and spin transfer with inserting flux $\theta_{\uparrow}=\theta,\theta_{\downarrow}=0$ in the cylinder: (a) bosonic systems on the $N_y=3$ cylinder at $\nu=2/3,U=10t,V=0$; (b) fermionic systems on the $N_y=4$ cylinder at $\nu=1/2,U=10t,V=20t$. Here the calculation is performed using infinite DMRG with keeping up to 1200 states.}
\end{figure}

To uncover the topological nature of two-component FQH states, we further calculate the charge pumping under the insertion of flux quanta on cylinder systems based on the newly developed adiabatic DMRG~\cite{Gong2014}
in connection to the quantized Hall conductance. It is expected that a quantized charge will be pumped from the right side to the left side by inserting a U(1) charge flux $\theta=0 \rightarrow 2\pi $. The net  transfer of the total charge from the right side to the left side is encoded by $Q(\theta)=N_{\uparrow}^{L}+N_{\downarrow}^{L}=tr[\widehat{\rho}_L(\theta)\widehat{Q}]$ ($\widehat{\rho}_L$ the reduced density matrix of the left part)~\cite{Zaletel2014}. In order to quantify the drag Hall conductance, we also define the spin transfer $\Delta S$ by $S(\theta)=N_{\uparrow}^{L}-N_{\downarrow}^{L}=tr[\widehat{\rho}_L(\theta)\widehat{S}]$ in analogy to the charge transfer.

As shown in Figs.~\ref{pump}(a) and~\ref{pump}(b), for bosons at $\nu=2/3$, a fractional charge $\Delta Q=Q(2\pi)-Q(0)\simeq0.33$ is pumped by threading one flux quanta with $\theta_{\uparrow}=\theta,\theta_{\downarrow}=0$ in one species of two-component gases, and a fractional charge $\Delta Q=Q(2\pi)-Q(0)\simeq 0.66 \simeq\nu$ would be pumped by threading one flux quanta with $\theta_{\uparrow}=\theta_{\downarrow}=\theta$ in both of two-component gases. For fermion at $\nu=1/2$, a fractional charge $\Delta Q=Q(2\pi)-Q(0)\simeq0.25$ is pumped by threading one flux quanta with $\theta_{\uparrow}=\theta,\theta_{\downarrow}=0$ in one species of two-component gases, and a fractional charge $\Delta Q=Q(2\pi)-Q(0)\simeq 0.50 \simeq\nu$ would be pumped by threading one flux quanta with $\theta_{\uparrow}=\theta_{\downarrow}=\theta$ in both of two-component gases.

The dynamical pumping process reveals the fractional statistics of the pumped quasiparticle. Based on these observations, we claim that the number of physically distinct stable quasiparticles is equal to the rank of the $\mathbf{K}$-matrix, and establish the pumping relationship that (i) by threading the flux $\theta_{\uparrow}=\theta,\theta_{\downarrow}=0$ from $\theta=0$ to $\theta=2\pi$
\begin{align}
  &\Delta Q=C_{\uparrow,\uparrow}+C_{\downarrow,\uparrow},\\
  &\Delta S=C_{\uparrow,\uparrow}-C_{\downarrow,\uparrow},
\end{align}
and (ii) by threading the flux $\theta_{\uparrow}=\theta_{\downarrow}=\theta$ from $\theta=0$ to $\theta=2\pi$
\begin{align}
  &\Delta Q=\sum_{\sigma,\sigma'}C_{\sigma,\sigma'},\\
  &\Delta S=0.
\end{align}

\subsection{Chiral edge spectrum}\label{edge}

\begin{figure}[t]
  \includegraphics[height=1.7in,width=3.4in]{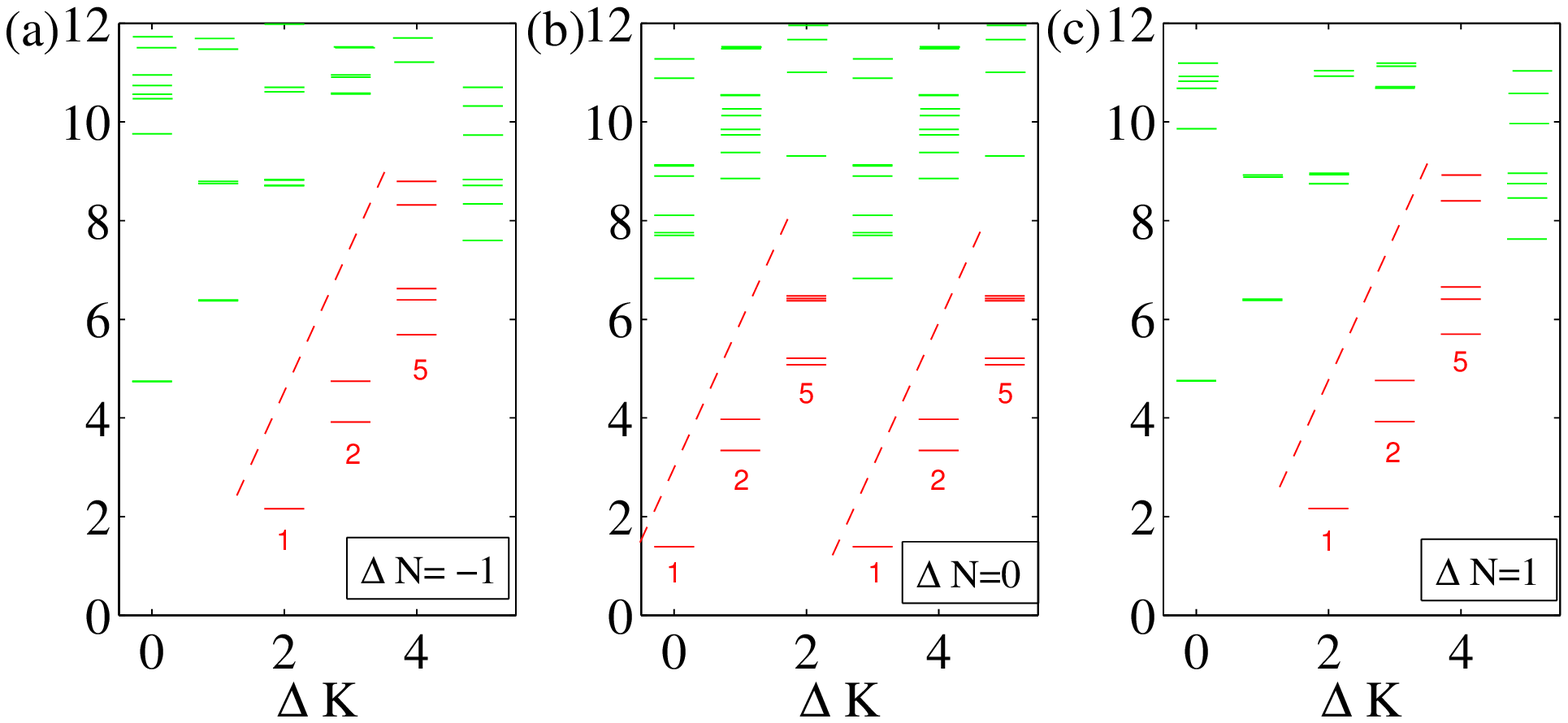}
  \caption{\label{es}(Color online) The momentum-resolved entanglement spectrum for bosonic systems on the $N_y=6$ cylinder of single layer square lattice with Chern number two at $\nu=1/3,U=10t,V=0$. In each tower, the horizontal axis shows the relative momentum $\Delta K=K_y-K_{y}^{0}$ (in units of $2\pi/N_y$) in the transverse direction of the corresponding eigenvectors of density matrix $\rho_L$. The numbers below the red dashed line label the nearly degenerating pattern for the low-lying ES with different momenta: $1,2,5,\cdots$}
\end{figure}

Another ``fingerprint'' of chiral topological order is the characteristic edge state usually described by the specific Luttinger liquid theory, which can be revealed through the low-lying entanglement spectrum (ES)~\cite{Li2008}. Here we examine the ES of these FQH states based on DMRG method. Due to its difficulty of DMRG convergence for two-component gases on the $N_y>6$ cylinder, in order to calculate the momentum-resolved entanglement spectrum, we stack the two-component topological checkerboard lattices into an equivalent single-layer square lattice with Chern number two, as engineered in Ref.~\cite{Yang2012}, where hardcore bosons at one-third filling host three-fold degenerate ground states reminiscent of Halperin (221) states. In such a construction, the original two degenerate edge modes in bilayer checkerboard lattices are differed by a momentum phase $\pi$ in the Brillouin zone of the single-layer square lattice.

For hardcore bosons, as shown in Figs.~\ref{es}(a-c), the two branches of low-lying bulk ES appear with the level counting $1,2,5,\cdots$, implying the gapless edge modes. Since $\mathbf{K}$ has two positive eigenvalues, the edge excitations would have two forward-moving branches in the same direction in the charge sector $\Delta N=0$ as shown in Fig.~\ref{es}(b), consistent with theoretical analysis of spin-singlet quantum Hall state in Refs.~\cite{Moore1997,Wu2012,Geraedts2015}.

\begin{figure}[t]
  \includegraphics[height=1.95in,width=3.2in]{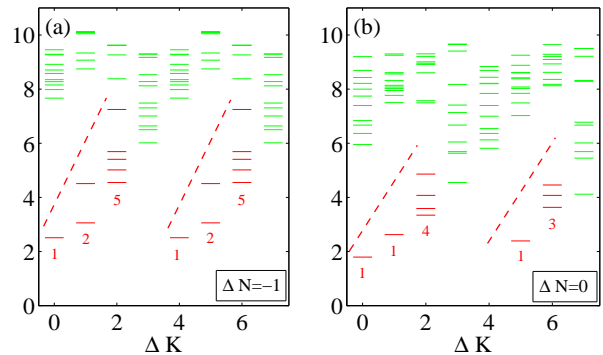}
  \caption{\label{es331}(Color online) The momentum-resolved entanglement spectrum for fermionic systems on the $N_y=8$ cylinder of single layer square lattice with Chern number two at $\nu=1/4,U=10t,V=20t$. In each tower, the horizontal axis shows the relative momentum $\Delta K=K_y-K_{y}^{0}$ (in units of $2\pi/N_y$) in the transverse direction of the corresponding eigenvectors of density matrix $\rho_L$. The numbers below the red dashed line label the nearly degenerating pattern for the low-lying ES with different momenta.}
\end{figure}

Similarly for fermions, the edge excitations would also have two forward-moving branches in the same direction. As shown in Figs.~\ref{es331}(a) and~\ref{es331}(b), for odd charge sector $\Delta N=-1$, two symmetric branches of ES appear with the same level counting $1,2,5,\cdots$, differing by a phase $\pi$ in the momentum. However, for even charge sector $\Delta N=0$, two asymmetric branches of ES appear with different level countings $1,1,4,\cdots$ and $0,1,3,\cdots$, differing by a phase $\pi$ in the momentum. Thus, without the $\pi$-phase shift, the total counting of two branches of ES would be (a) $2,4,10,\cdots$ for odd charge sector $\Delta N=-1$ and (b) $1,2,7,\cdots$ for even charge sector $\Delta N=0$, which are consistent with the analysis of the counting of the root configurations of the bilayer Halperin (331) edge excitations in the lowest Landau level at half filling~\cite{Zhu2016}.

\section{Integer quantum Hall state}\label{iqh}
In this section, we consider the ground states at integer filling $\nu=2N/N_s=2$ for two-component fermions ($N=N_{\uparrow}+N_{\downarrow}$), namely, one fermion per site. In noninteracting cases $U=0,V=0$, the system is a topological Chern insulator with quantized Hall conductivity two. In the large repulsive case $U\gg t$, it would be a trivial antiferromagnetic Mott insulator as expected. Several mean field studies of Haldane-honeycomb model indicate a topologically nontrivial N\'{e}el antiferromagnetic insulating phase in the intermediate interaction regime~\cite{Kou2011,Zheng2015,Arun2016}, supported by quantum cluster methods~\cite{Wu2016}. However recent dynamical cluster approximation
have revealed a first order transition from $C_q=2$ Chern insulator into a trivial Mott insulator, with no evidence of topological antiferromagnetic insulator as an intermediate phase~\cite{Vanhala2016,Troyer2016}. In the following discussion, we take the typical example of $\pi$-flux checkerboard lattice, and investigate the nature of the interaction-driven transition using both ED and DMRG calculations.

\begin{figure}[t]
  \includegraphics[height=2.9in,width=3.4in]{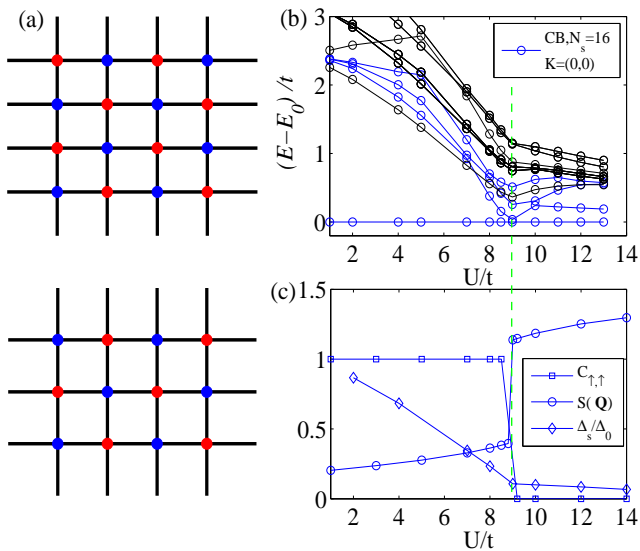}
  \caption{\label{crossover}(Color online) Numerical results for the y-direction spectral flow of fermionic systems : (a) Two different checkerboard lattice geometries considered in our calculation; (b) The low energy spectra as a function of onsite repulsion $U$ for system $N_s=16,N_{\uparrow}=N_{\downarrow}=8$ at $\nu=2$; (c) The manybody Chern number $C_{\uparrow,\uparrow}$ and antiferromagnetic spin structure factor $S(\mathbf{Q})$ of the lowest ground state, and the spin excitation gap $\Delta_s$ as a function of onsite repulsion $U$. Here the parameters $V=0,t'=0.3t,t''=0$.}
\end{figure}

For small system sizes, we present a ED diagnosis of quantum phase transition of Fermi-Hubbard model from weak interactions to strong interactions in $\pi$-flux checkerboard lattice. Our current system size  limit for ED calculation is
 $N_s=16,N_{\uparrow}=N_{\downarrow}=8$. For simplicity, we take $t'=0.3t,t''=0$. The lattice geometry is indicated in Fig.~\ref{crossover}(a), with two different lattice sizes $N_s=16,12$. In Fig.~\ref{crossover}(b), we plot its low energy evolution as onsite repulsion $U$ increases. For weak interactions, there always exists a stable unique ground state with a large gap separated from higher levels. When one flux quanta is inserted, it would evolves back to  itself. With twisted boundary conditions as described in Sec.~\ref{kmatrix}, we obtain the $\mathbf{K}$-matrix as a unit matrix
\begin{align}
  \mathbf{K}=\mathbf{C}=\begin{pmatrix}
1 & 0\\
0 & 1\\
\end{pmatrix}.
\end{align}
When $U$ increases further, this ground state undergoes an energy level crossing with other levels around $U_c\simeq9t$. In order to clarify the nature of the phase transition, we calculate its Chern numbers $C_{\uparrow,\uparrow},C_{\downarrow,\downarrow}$ and antiferromagnetic spin structure factor
\begin{align}
   S(\mathbf{Q})=\frac{1}{N_s}\sum_{\rr,\rr'}e^{i\mathbf{Q}\cdot(\rr-\rr')} \langle S^{\rr}_zS^{\rr'}_z \rangle,
\end{align}
where $\mathbf{Q}=(\pi,\pi),S_z=(n_{\uparrow}-n_{\downarrow})/2$. As indicated in Fig.~\ref{crossover}(c), both $C_{\uparrow,\uparrow}=C_{\downarrow,\downarrow}$ and $S(\mathbf{Q})$ experience a sudden dramatic jump as the interaction $U$ increases across the critical threshold $U_c$, demonstrating the discontinuous first-order transition. In the Mott regime, the antiferromagnetic order dominates and we obtain the charge Hall conductance $C_q=C_{\uparrow,\uparrow}+C_{\downarrow,\downarrow}=0$, as expected. Meanwhile, the spin gap $\Delta_s=E_0(S_z=1)-E_0(S_z=0)$ tends to collapse (here a finite small value is limited by finite size effects).

\begin{figure}[t]
  \includegraphics[height=3.0in,width=3.0in]{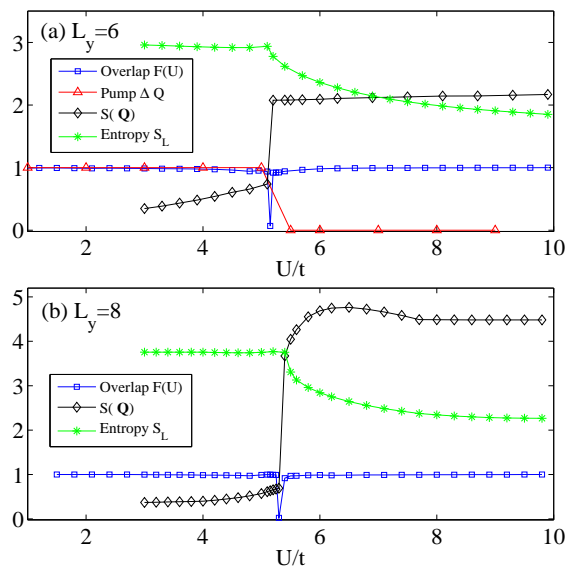}
  \caption{\label{dmrg}(Color online) Numerical DMRG results on a cylinder with finite width $L_y=2\times N_y$ and fixed length $L_x=18$ for $N_{\uparrow}=N_{\downarrow}=L_x\times N_y,\nu=2$. The evolutions of physical quantities vs onsite repulsion $U$, including the absolute wavefunction overlap $F(U)$, charge pumping $\Delta Q$, spin structure factor $S(\mathbf{Q})$, and entanglement entropy $S_{L}$ are shown for (a) $L_y=6,N_y=3$ and (b) $L_y=8,N_y=4$, respectively. The first-order transition is characterized by the discontinuous behavior of these physical quantities at the transition point. Here the parameters $V=0,t'=0.3t,t''=0$, and the maximally kept number of states 2020.}
\end{figure}

For larger system sizes, we exploit an unbiased DMRG approach to study this phase transition from four different aspects, using a cylinder geometry up to a maximum width $L_y=8$ ($N_y=4$). As shown in Figs.~\ref{dmrg}(a) and~\ref{dmrg}(b), first, we calculate the absolute wavefunction overlap $F(U)=|\langle\psi(U)|\psi(U+\delta U)\rangle|$ where $\delta U$ is a small finite value (here we take $\delta U$ as small as $0.05t$ near the transition region). For $U<U_c$ (or $U>U_c$), $F(U)$ has a large value close to 1, indicating the same structure of wavefunctions. When $U$ approaches a critical value $U_c$~\cite{noteU}, $F(U)$ suddenly drops down to a very small value close to zero, separating two different phases. Second, we calculate the charge pumping by inserting one flux $\theta_{\uparrow}=\theta,\theta_{\downarrow}=0$ from $\theta=0$ to $\theta=2\pi$, and obtain : (i) $\Delta Q=1,\Delta S=1$ in weak interacting regime $U<U_c$; (ii) $\Delta Q=0,\Delta S=0$ in strongly interacting regime $U>U_c$. Third, the spin structure factor $S(\mathbf{Q})$ is also calculated, and it exhibits a discontinuous jump near the transition point, similar to our ED analysis.

Finally, we uncover its entanglement signatures of quantum Hall transitions~\cite{Gu2004,Zozulya2009}. We partition the system into two halves at the cylinder center and trace out the right part to obtain the entanglement spectrum $\xi_i$ (the eigenvalues of reduced density matrix of the left part) and entanglement entropy $S_L=-\sum_i\xi_i\ln\xi_i$. As shown in Figs.~\ref{dmrg}(a) and~\ref{dmrg}(b), for $U<U_c$, $S_L$ is almost a constant for a given system, indicating that the same topological properties as integer quantum Hall state. $S_L$ starts to drop at $U\simeq U_c$, with a discontinuous first-order derivative $\partial S_L/\partial U$ due to the change of topology~\cite{note}. By comparing the complete spectra of reduced density matrices between two ground states, the first-order transition can be extracted from the discontinuous jump of majorization~\cite{Zozulya2009}. Recently, a diagnostic of phase transition driven by disorder from quantum Hall states to an insulator via entanglement entropy is also examined~\cite{Liu2016}.

The above pictures from both ED and DMRG calculations indicate a first-order phase transition directly from a $C_q=2$ integer quantum Hall state to a Mott insulator driven by onsite repulsion. We did not observe any evidence of topological Neel antiferromagnetic insulating phase in the intermediate interaction regime, which is consistent with recent studies on the Haldane-honeycomb lattice from different methods~\cite{Vanhala2016,Troyer2016}. Further, with the inclusion of next-next-nearest-neighbor hopping $t''=-0.2t$, we numerically arrive at the same phase transition nature as above. For the future work, it would be interesting to consider the effects of the band flatness modulated by next-nearest-neighbor hopping $t'$ and next-next-nearest-neighbor hopping $t''$ on the quantum phase diagram and phase transition.

\section{Summary and Discussions}\label{summary}

In summary, we show that two-component hardcore bosons and fermions in topological lattice models could host Halperin FQH states at a partial filling of the lowest Chern band, with fractional topological properties characterized by $\mathbf{K}$-matrix, including the degeneracy, fractional quantized Hall conductance and chiral edge modes. The role of onsite interspecies repulsion on integer quantum Hall state on the $\pi$-flux checkerboard topological lattice is examined, and shown to lead to a first-order phase transition from a $C_q=2$ integer quantum Hall state to a trivial Mott insulator for two-component fermions at integer filling factor $\nu=2$. Another interesting issue for two-component hardcore bosons at $\nu=2$ filling, but not discussed here, is related to the possibility for a bosonic integer quantum Hall state~\cite{Senthil2013,Furukawa2013,Regnault2013,YHWu2013,Ye2013,Sterdyniak2015,He2015,Zeng2016}, which is left for future study.

\begin{acknowledgements}
This work is supported by the U.S. Department of Energy, Office of Basic Energy Sciences under Grants No. DE-FG02-06ER46305 (T.S.Z, and D.N.S.), and and by U.S. Department of Energy{'}s National Nuclear Security Administration through Los Alamos National Laboratory LDRD Program (W.Z.).
\end{acknowledgements}


\begin{thebibliography}{99}
\bibitem{noteU} The different critical values obtained by ED and DMRG are possibly due to finite-size and boundary effects.
\end{thebibliography}
\end{document}